\documentstyle[12pt]{article}

\def\be{\begin{equation}}
\def\ee{\end{equation}}
\def\ba{\begin{array}{c}}
\def\ea{\end{array}}

\def\ben{$$}
\def\een{$$}
\begin{document}

\titlepage
\vspace*{4cm}

\begin{center}{\Large \bf
New set of exactly solvable complex potentials giving the real
energies }\end{center}

\vspace{5mm}

\begin{center}
Miloslav Znojil
\vspace{3mm}

\'{U}stav jadern\'e fyziky AV \v{C}R, 250 68 \v{R}e\v{z},
Czech Republic\\

e-mail: znojil@ujf.cas.cz

\end{center}

\vspace{5mm}

\section*{Abstract}

We deform the real potential $V(x)$ of P\"{o}schl and Teller by a
shift $\varepsilon \in (0, \pi/2)$ of $x$ in imaginary direction.
We show that the new model $V(x)=F/\sinh^2 (x-i\,\varepsilon) +
G/\cosh^2 (x-i\,\varepsilon)$ remains exactly solvable. Its bound
states are constructed in closed form. Wave functions are complex
and proportional to Jacobi polynomials. Some of them diverge in
the limit $\varepsilon \to 0$ or $\varepsilon \to \pi/2$. In
contrast, all their energies prove real and
$\varepsilon-$independent. In this sense the loss of Hermiticity
of our family of Hamiltonians seems well counter-balanced by their
accidental ${\cal PT}$ symmetry.

\vspace{9mm}

\noindent
 PACS 03.65.Ge,
03.65.Fd



\newpage

\noindent Among all the exactly solvable models in quantum
mechanics the one-di\-men\-si\-o\-nal Schr\"{o}dinger equation
 \be
\left (-\,\frac{d^2}{dr^2} + V(r) \right )
 \, \psi(r) = E  \, \psi(r), \ \ \ \ \ \psi(\pm \infty) = 0
\label{SE}  \ee with one of the most elementary bell-shaped
potentials $V^{(bs)}(r) = G/{\cosh^2 r}$ is particularly useful.
Its applications range from the analyses of stability and
quantization of solitons \cite{Bullough} to phenomenological
studies in atomic and molecular physics \cite{physaa}, chemistry
\cite{physbb}, biophysics \cite{physcc} and astrophysics
\cite{physdd}.  Its appeal involves the solvability by different
methods \cite{Levai} as well as a remarkable role in the
scattering \cite{Newton}. Its bound-state wave functions
represented by Jacobi polynomials offer one of the most elementary
illustrations of properties of the so called shape invariant
systems \cite{Cooper}. The force $V^{(bs)}(r)$ is encountered in
the so called ${\cal PT}$ symmetric quantum mechanics
\cite{Bender} where it appears as a Hermitian super-partner of a
complex ``scarf" model \cite{RKRC}.

Curiously enough, it is not too difficult to extend the exact
solvability of the potential $V^{(bs)}(r) $ to all its ``spiked"
(often called P\"{o}schl-Teller \cite{Poeschl}) shape invariant
generalizations
 \be
V^{(PT)}(r)
= -\frac{A(A+1)}{\cosh^2 r} +\frac{B(B-1)}{\sinh^2 r}.
\label{sPTP}
  \ee
Unfortunately, as far as the one-dimensional Schr\"{o}dinger eq.
(\ref{SE}) becomes too singular at $B(B-1) \neq 0$ the more
general force $V^{(PT)}(r) $ must be confined to the semi-axis,
i.e., in most cases, to the $s$ wave in three dimensions with
coordinates $ r \in (0, \infty)$. This makes the ``improved"
P\"{o}schl-Teller model (\ref{sPTP}) much less useful in practice
since its higher partial waves are not solvable. The impossibility
of using eq. (\ref{sPTP}) in three dimensions (or on the whole
axis in one dimension at least) is felt unfortunate because the
singular potentials themselves are frequently needed in methodical
considerations \cite{Klauder} and in perturbation theory
\cite{Harrell}. They are encountered in phenomenological models
\cite{Sotona} and in explicit computations \cite{Hall} but not too
many of them are solvable \cite{Mathieu}. This was a strong
motivation of our present brief note on eq. (\ref{SE}) +
(\ref{sPTP}).

We feel inspired by the pioneering letter \cite{BB} where Bender
and Boettcher introduced the so called ${\cal PT}$ symmetry
(meaning the commutativity of a complex Hamiltonian with the
product of parity ${\cal P}$ and time reversal ${\cal T}$). They
proposed its use as a possible source of the reality of spectra
for non-Hermitian Hamiltonian operators. For illustration they
employed the harmonic oscillator $V^{(HO)}(r) = r^2$ with the
complex downward shift of its axis of coordinates,
 \be
r = x-i\varepsilon, \ \ \ \ \ \  \ x \in (-\infty,\infty).
\label{shov}  \ee The ${\cal PT}$ symmetry of their model
$V^{(BB)}(x)=V^{(HO)}(x-ic) = x^2 - 2icx - c^2$ means its
invariance with respect to the simultaneous reflection $x \to -x$
and complex conjugation $i \to -i$.  Their example inspired their
general hypothesis that the ${\cal PT}$ symmetry could by itself
imply the reality of spectrum in some non-Hermitian models
\cite{Bender}.

Various other complex interactions have been tested and studied
within this framework \cite{review}. In particular, the
three-dimensional ${\cal PT}$ symmetric harmonic oscillator of
ref. \cite{ptho} offers us another key idea. The same shift
(\ref{shov}) has been employed there as a source of a {\em
regularization} of the strongly singular centrifugal term. As long
as $1/(x- i\varepsilon)^2 = (x+i\varepsilon)^2 /
(x^2+\varepsilon^2)^2 $ at any $\varepsilon \neq 0$, this term
remains nicely bounded in a way which is uniform with respect to
$x$. Without any difficulties one may work with $V^{(RHO)}(x) =
r^2(x) +\ell(\ell+ 1)/ r^2(x)$ on the whole real line of $x$. In
what follows the same idea will be applied to the regularized
P\"{o}schl-Teller-like potential
 \ben V^{(RPT)}(x) = V^{(PT)}(x -
i\varepsilon), \ \ \ \ \ \ \ 0 < \varepsilon <\pi/2.
  \een
This potential is a simple function of the L\'{e}vai's
\cite{Levai} variable $g(r)=\cosh 2r$. As long as $g(x -
i\,\varepsilon) = \cosh 2x\,\cos 2\varepsilon - i\,\sinh 2x\,\sin
2\varepsilon$, the new force is ${\cal PT}$ symmetric on the real
line of $x \in (-\infty, \infty)$,
  \ben V^{(RPT)}(-x)= [V^{(RPT)}(x)]^*.
  \een
Due to the estimates $|\sinh^2(x-i\varepsilon)|^2 = \sinh^2 x
\cos^2 \varepsilon +\cosh^2 x \sin^2 \varepsilon = \sinh^2 x
+\sin^2 \varepsilon$ and $|\cosh^2(x-i\varepsilon)|^2 = \sinh^2 x
+\cos^2 \varepsilon$ the regularity of $V^{(RPT)}(x)$ is
guaranteed for all its parameters $\varepsilon \in (0, \pi/2)$.

In a way paralleling the three-dimensional oscillator the mere
analytic continuation of the $s-$wave bound states does not give
the complete solution.  One must return to the original
differential equation (\ref{SE}). There we may conveniently fix $A
+1/2=\alpha>0$ and $B-1/2=  \beta>0$ and write
 \be
\left (-\,\frac{d^2}{dx^2} +
\frac{\beta^2-1/4}{\sinh^2 r(x)}
 -\frac{\alpha^2-1/4}{\cosh^2 r(x)}  \right )
 \, \psi(x) = E  \, \psi(x), \ \ \ \ \ \ r(x)= x-i\varepsilon.
\label{SEb}
 \ee
This is the Gauss differential equation
 \be
z(1+z)\,\varphi''(z) +[c+(a+b+1)z]\,\varphi'(z)
+ab\,\varphi(z)=0
\label{gauss}
 \ee
in the new variables
 \ben
\psi(x) = z^\mu(1+z)^\nu\varphi(z),
\ \ \ \ \ \ \ \ z = \sinh^2r(x)
 \een
using the suitable re-parameterizations
 \ben
\alpha^2=(2\nu-1/2)^2,
\ \ \ \ \ \ \ \ \
 \beta^2=(2\mu-1/2)^2,
\ \ \ \ \ \ \ \ \
 \een
 \ben
2\mu+1/2=c, \ \ \ \ \
2\mu+2\nu=a+b, \ \ \ \ \
E= -(a-b)^2.
 \een
In the new notation we have the wave functions
 \be
\psi(x) = \sinh^{\tau \beta+1/2}[r(x)]
 \cosh^{\sigma\alpha+1/2}[r(x)]\,\varphi[z(x)]
\label{formula}
  \ee
with the sign ambiguities $\tau = \pm 1$ and $\sigma=\pm 1$ in
$2\mu=\tau \beta+1/2$ and $2\nu=\sigma\alpha+1/2$. This formula
contains the general solution of hypergeometric eq. (\ref{gauss}),
 \be
\varphi(z) = C_1\ _2F_1(a,b;c;-z) + C_2z^{1-c}\
_2F_1(a+1-c,b+1-c;2-c;-z). \label{gensol}
  \ee
The solution should obey the complex version of the
Sturm-Liouville oscillation theorem \cite{Hille}.  In the case of
the discrete spectra this means that we have to demand the
termination of our infinite hypergeometric series. This suppresses
an asymptotic growth of $\psi(x)$ at $x\to\pm\infty$.

In a deeper analysis let us first put $C_2=0$. We may satisfy the
termination condition by the non-positive integer choice of
$b=-N$.  This implies that $a=N+1 +\sigma\alpha +\tau  \beta$ is
real and that our wave function may be made asymptotically
(exponentially) vanishing under certain conditions. Inspection of
the formula (\ref{formula}) recovers that the boundary condition
$\psi(\pm \infty) = 0$ will be satisfied if and only if
  \ben 1 \leq 2N+1\leq 2N_{max}+1<-\sigma
\alpha - \tau  \beta.
 \een The closed Jacobi polynomial
representation of the wave functions follows easily,
  \ben
\varphi[z(x)] =C_1\ \frac{N!\Gamma(1+\tau \beta)}{\Gamma(N+1+\tau
\beta)} \ P_N^{(\tau \beta,\sigma\alpha)}[\cosh 2r(x)].
  \een
The final insertions of parameters define the spectrum of
energies,
 \be
E=-( 2N+1+\sigma \alpha + \tau  \beta)^2 < 0. \label{energy}
 \ee
Now we have to return to eq.  (\ref{gensol}) once more. A careful
analysis of the other possibility $C_1=0$ does not recover
anything new.  The same solution is obtained, with $\tau$ replaced
by $-\tau$.  We may keep $C_2=0$ and mark the two independent
solutions by the sign $\tau$. Once we define the maximal integers
$N_{max}^{(\sigma,\tau)}$ which are compatible with the inequality
 \be
2N_{max}^{(\sigma,\tau)}+1< -\sigma\alpha-\tau \beta \label{maxes}
 \ee
we get the constraint $N \leq N_{max}^{(\sigma,\tau)}$. The set of
our main quantum numbers is finite.

Let us now compare our final result (\ref{energy}) with the known
$\varepsilon=0$ formulae for $s$ waves \cite{Levai}. An additional
physical boundary condition must be imposed in the latter singular
limit \cite{conditab}.  This condition fixes the unique pair
$\sigma = -1$ and $\tau = +1$. Thus, the set of the $s-$wave
energy levels $E_N$ is not empty if and only if $\alpha- \beta>
1$. In contrast, all our $\varepsilon > 0$ potentials acquire a
uniform bound $|V^{(RPT}(x)| < const < \infty$. Due to their
regularity, no additional constraint is needed. Our new spectrum
$E^{(\sigma,\tau)}_N$ becomes richer. For the sufficiently strong
couplings it proves composed of the three separate parts,
 \ben
E^{(-,-)}_N< 0, \ \ \ \ \ 0 \leq N \leq N_{max}^{(-,-)}, \ \ \ \ \
\ \alpha+ \beta > 1,
 \een
 \be
E^{(-,+)}_N<0, \ \ \ \ \ \ 0 \leq  N \leq N_{max}^{(-,+)}, \ \ \ \
\ \ \alpha> \beta + 1,
 \ee
 \ben
E^{(+,-)}_N<0, \ \ \ \ \ \ 0 \leq  N \leq N_{max}^{(+,-)},
 \ \ \ \ \ \
\beta > \alpha+1.
 \een
The former one is non-empty at $ A + B > 1$ (with our above
separate conventions $A > -1/2$ and $B > 1/2$). Concerning the
latter two alternative sets, they may exist either at  $A> B$ or
at $B > A+2$, respectively. We may summarize that in a parallel to
the ${\cal PT}$ symmetrized harmonic oscillator of ref.
\cite{ptho} we have the $N_{max}^{(-,+)}+1$ quasi-odd or
``perturbed", analytically continued $s-$wave states (with a nodal
zero near the origin) complemented by certain additional
solutions.

In the first failure of a complete analogy the number
$N^{(-,-)}_{max}+1$ of our quasi-even states proves systematically
higher than $N^{(-,+)}_{max}+1$, especially at the larger
``repulsion" $ \beta \gg 1$. This is a certain paradox,
strengthened by the existence of another quasi-odd family which
behaves very non-perturbatively.  Its members (with the ground
state $\psi_0^{(+,-)}(x) =\cosh^{A+1} [r(x)]\sinh^{1-B} [r(x)]$
etc) do not seem to have any $s-$wave analogue. They are formed at
the prevalent repulsion $B>A+2$ which is even more
counter-intuitive. The exact solvability of our example enables us
to understand this apparent paradox clearly. In a way
characteristic for many ${\cal PT}$ symmetric systems some of the
states are bound by an antisymmetric imaginary well. The whole
history of the ${\cal PT}$ symmetric models starts from the purely
imaginary cubic force \cite{Bessis} after all. A successful
description of its perturbative forms $V(x) = \omega x^2+i\lambda
\,x^3$ is not so enigmatic \cite{Graffi}, especially due to its
analogies with the real and symmetric $V(x) = \omega x^2+
\lambda\,x^4$ \cite{Alvarez}. The similar mechanism creates the
states with $(\sigma,\tau)=(+,-)$ in the present example. A
significant novelty of our new model $V^{(RPT)}(x)$ lies in the
dominance of its imaginary component {\em at the short distances},
$x \approx 0$. Indeed, we may expand our force to the first order
in the small $\varepsilon>0$. This gives the approximation
 \be
\frac{1}{\sinh^2(x-i\varepsilon)}=
\frac{\sinh^2(x+i\varepsilon)}{(\sinh^2x + \sin^2\varepsilon)^2}
=\frac{1}{\sinh^2x}+2i\varepsilon \frac{\cosh x}{\sinh^3 x}
+{\cal O}(\varepsilon^2).
 \label{sini}
 \ee
We see immediately the clear prevalence of the imaginary part at
the short distances, especially at all the negligible $A = {\cal
O}(\varepsilon^2)$.

An alternative approach to the above paradox may be mediated by a
sudden transition from the domain of a small $\varepsilon \approx
0$ to the opposite extreme with $\varepsilon \approx \pi/2$. This
is a shift which changes $\cosh x$ into $\sinh x$ and vice versa.
It intertwines the role of $\alpha$ and $ \beta$ as a strength of
the smooth attraction and of the singular repulsion, respectively.
The perturbative/non-perturbative interpretation of both our
quasi-odd subsets of states becomes mutually interchanged near
both the extremes of $\varepsilon$.

The dominant part (\ref{sini}) of our present model leaves its
asymptotics comparatively irrelevant. In contrast to many other
${\cal PT}$ symmetric models as available in the current
literature our potential vanishes asymptotically,
 \ben V^{(RPT)}(x) \to 0, \ \ \ \ \ \ \ \ x \to \pm \infty.
  \een
An introduction and analysis of continuous spectra in the ${\cal
PT}$ symmetric quantum mechanics seems rendered possible at
positive energies. This question will be left open here. In the
same spirit of a concluding remark we may also touch the problem
of the possible breakdown of the ${\cal PT}$ symmetry. This has
recently been studied on the background of the supersymmetric
quantum mechanics \cite{Canata}. In our present solvable example
the violation of the ${\cal PT}$ symmetry is easily mimicked by
the complex choice of the couplings $\alpha$ and $ \beta$. Due to
our closed formulae the energies will still stay real, provided
only that ${\rm Im}\ (\sigma\alpha+ \tau \beta)=0$. Unfortunately,
the questions of this type lie already beyond the scope of our
present short communication.

\end{document}